\documentclass[twocolumn,showpacs,preprintnumbers,amsmath,amssymb]{revtex4}
\usepackage{graphicx}
\usepackage{bm}

\begin{document}
\preprint{...}
\title{Efficient readout of micromechanical resonator arrays in ambient conditions}
\author{W.J. Venstra} \email{w.j.venstra@tudelft.nl}
\author{H.S.J. van der Zant}\email{h.s.j.vanderzant@tudelft.nl}
\affiliation{Kavli Institute of Nanoscience, Delft University of Technology, Lorentzweg 1, 2628CJ Delft, The Netherlands}%
\date{\today}
\begin{abstract}
We present a method for efficient spectral readout of mechanical
resonator arrays in dissipative environments. Magnetomotive drive
and detection is used to drive double clamped resonators in the
nonlinear regime. Resonators with almost identical resonance
frequencies can be tracked individually by sweeping the drive
power. Measurements are performed at room temperature and
atmospheric pressure. These conditions enable application in high
throughput resonant sensor arrays.
\end{abstract}

\maketitle

Micro- and nanometer scale mechanical resonators are widely
considered as mass and force sensors. Ad- or desorption of
molecules by the resonator are detected as slight changes in
resonance frequency. Several methods exist to measure the
resonance frequency of a single resonator
\cite{Karabacak06,Li07,Dohn06, Postma05}. Small arrays of static
mechanical sensors be scanned optically \cite{Mertens05} or the
reflection of the array as a whole may be analyzed \cite{Yue08}.
Spectral readout of small-scale resonant arrays has been reported
in vacuum \cite{Ekinci07}. In this letter we present an efficient
method to readout resonators with closely spaced resonance
frequencies in dissipative environments. We show that the number
of resonant sensors which can be operated in a given bandwidth is
not limited by the Q factor. Using a magnetomotive setup we can
discriminate mechanical resonators with almost identical
frequencies in ambient conditions.\\
\indent\indent To track two individual resonators with quality
factor $Q$ in a frequency spectrum, their resonance frequencies
$f_i$ should be separated by at least $\Delta f=f_i/Q$.
Simultaneous operation of $n$ resonators requires a minimum
bandwidth on the order of $\Delta f=f_1(1+1/Q)^n$. Here $f_1$ is
the resonance frequency of the slowest resonator. In vacuum, the
Q-factor of a mechanical resonator can be of the order $10^4$
\cite{Ekinci07}, and readout of large arrays of resonators should
be feasible. At atmospheric pressure however, Q-factors usually
drop below $10^2$ because of viscous drag \cite{Sader06JAP}. This
limits the number of resonators within a practical bandwidth to a
few hundred at most. Large-scale arrays therefore require either
multiple detectors or separated drive circuits, and the complexity
of such systems rapidly increases with the number of resonators.\\
\indent\indent To address this problem, we propose to readout
arrays of resonators with closely spaced center frequencies by
sweeping the drive power from low to high values at constant
frequency. Once the nonlinear regime is accessed, individual
resonators are marked by instantaneous transitions in the phase
and amplitude response. When each nonlinear resonator is used as a
sensing element \cite{Greywall05}, the locations of these
characteristic amplitudes are affected by ad- or desorption
induced mass or stress change to the concerning resonator. We
conduct experiments on arrays of double clamped resonators at room
temperature and atmospheric pressure. We experimentally show that
in a 2-D array of $6$ resonators with closely spaced resonance
frequencies, individual resonators can be discriminated using a
single drive and detector unit. Calculations
confirm the experimental findings.\\
\indent\indent Arrays of double clamped mechanical resonators are
fabricated out of $100$ nm thick low-pressure chemical vapor
deposited (LPCVD) silicon nitride by electron beam lithography.
All resonators have the same dimensions $200\mathrm{\mu
m}\times15\mathrm{\mu m}$. Slight variations in center frequencies
of the resonators are expected as a result of variations in
residual stress \cite{Gardeniers96}. Figure 1(a) shows one of the
resonators before metallization. A layer of chromium and gold is
evaporated on top to enable magnetomotive drive and detection. The
resonators share a support on one side, while supports at the
other side are separated. This allows the measurement of both the
individual and the collective responses. \\
\begin{figure}[b]
\includegraphics [width=85mm] {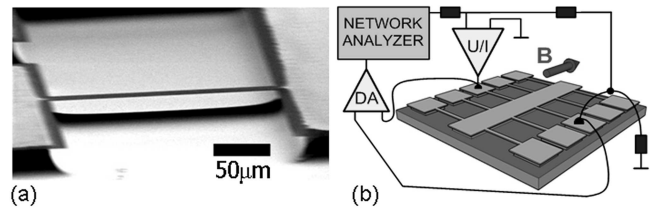}
\caption{(a) Scanning electron micrograph of a single resonator.
(b) Magnetomotive measurement setup. The current source is formed
by the amplifier U/I and the resistor network.\\}
\end{figure}
\indent\indent Figure 1(b) shows the measurement setup
schematically. The RF voltage from a network analyzer is converted
to a current which drives the resonators (two in series depicted).
A magnetic field of $1.9$ $\mathrm T$ is generated by a Halbach
array \cite{Halbach83} constructed from NdFeB permanent magnets.
The resulting voltage is amplified and measured by the network
analyzer. This measurement setup can be used at room temperature
and atmospheric pressure, in contrast to earlier magnetomotive
experiments in vacuum and/or at cryogenic temperatures. As a
result of viscous drag, the Q-factors in our experiments are a
factor of $10$ to $100$ lower than in those
experiments.\\
\begin{figure} [t]
\includegraphics [width=85mm] {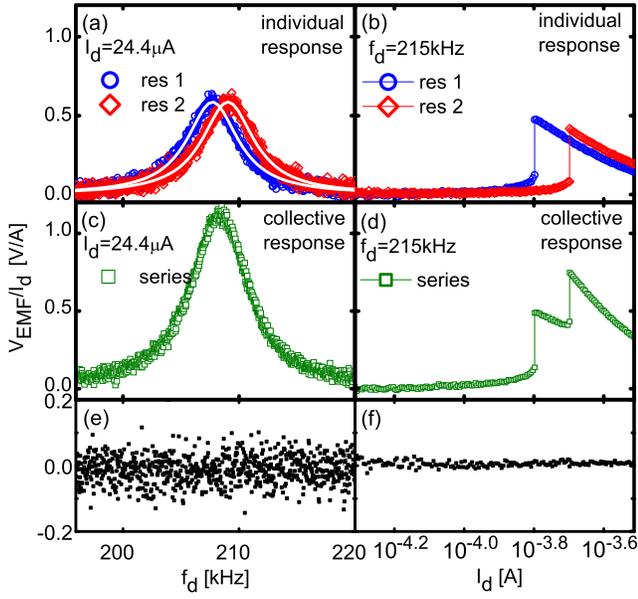}
\caption{Magnetomotive measurements in ambient environment. (a)
Individual linear frequency response of two resonators with almost
identical resonance frequencies. (b) The same resonators driven
nonlinear by sweeping drive power at $215 \mathrm{kHz}$. (c,d)
Collective response of the resonators connected in series in
linear (c) and nonlinear (d) regimes. (e,f) show the difference
between the calculated sum of the individual responses and the
measured collective response in the linear (e) and the nonlinear
(f) regime.}
\end{figure}
\indent\indent Figure 2(a) shows the linear response of two
resonators with almost identical center frequencies. The voltage
drop due to the resistance of the resonator, bond wires etc. has
been subtracted. The resonance frequencies and Q-factors are
determined by fitting Lorentzian functions (shown in the figure)
to the measured data. The resonance frequencies of these devices
were $207.72 \mathrm{kHz}$ and $209.05 \mathrm{kHz}$, and the
Q-factor is $39$ for both resonators. The corresponding bandwidth
is  $5.4 \mathrm{kHz}$, whereas the difference in resonance
frequency is $1.3 \mathrm{kHz}$. Now the resonators are connected
in series and the collective response is measured. Figure 2(b)
shows the result for the same drive conditions as in (a).
Discrimination between the two resonators is impossible as the
resonator bandwidth is more than four times the difference in
center frequency. We now individually drive the same resonators at
increasing power at a constant frequency of $215 \mathrm{kHz}$.
Fig. 2(d) shows the result: a steep transition in the resonator
amplitude marks the characteristic drive amplitude at this
frequency. When the collective response is measured at strong
driving, as in (d), the two resonators are easily distinguished.\\
\indent\indent In absence of interaction, the response of multiple
resonators is just the sum of the individual responses. To verify
this, the difference between the sum of the individual responses
of (a) and (d) and the collective response in (b) and (e) is
plotted in Fig.2 (c) and (f). No amplitude dependency is found,
which indicates that the coupling is weak. Note that a high
driving power results in a notably lower noise
level in the nonlinear driven system \cite{Cleland02}.\\
\begin{figure}[t]
\includegraphics [width=85mm] {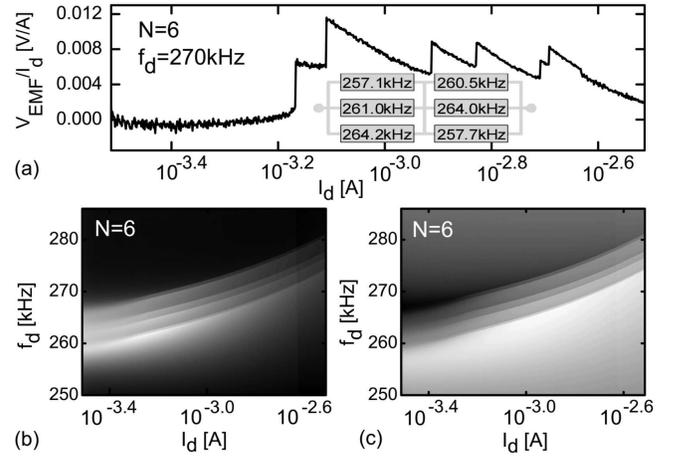}
\caption{Measurements on a 2-D array of 6 resonators. (a)
Resonator response while sweeping the drive current at $f=270
\mathrm{kHz}$. The inset shows the arrangement of the resonators
and their linear resonance frequencies. (b) and (c) show amplitude
and phase response of the array, driven at different frequencies
$f_d$.}
\end{figure}
\indent\indent We now turn to a 2-D array of $6$ resonators, for
which the arrangement and the measured linear resonance
frequencies are depicted in the inset of Fig.3(a). In this
experiment, the bandwidth of the individual resonators was
$5\pm0.1$ $\mathrm{kHz}$, which corresponds to a Q-factor of
approximately $50$. \\
\indent\indent Figure 3(a) shows the nonlinear amplitude response
at $f_{d}=270\mathrm{kHz}$; clearly 6 resonators can be
distinguished. In Fig.3(b) and (c), sweeps at fixed drive
frequencies ranging from $250\mathrm{kHz}$ to $285\mathrm{kHz}$
are plotted. Six lines marking characteristic jumps in resonator
amplitude and phase remain equidistant. Indeed, over the nonlinear
regime, the total bandwidth covering the $6$ resonators changes by
less than $60\mathrm{Hz}$ in this measurement. When sweeping
frequency from low to high values at constant drive current (not
shown), we found strong coupling between the resonators which
causes multiple resonators to collapse simultaneously at high
driving power. Strong coupling makes the system useless as an
array of independent sensors.\\
\indent\indent As for the differences in the center frequencies of
the resonators, we note that in LPCVD silicon nitride residual
stress variations of the order of $10\%$ across a $2'$ wafer are
not unusual \cite{Gardeniers96}. Using this number, the residual
stress gradient would be on the order of $100$ $\mathrm{Pa / \mu
m}$. Given the distance between the resonators (at least $300
\mathrm{\mu m}$) the difference in residual stress is on the order
of $10^4$ $\mathrm{Pa}$ which results in frequency differences on
the order of $100 \mathrm {Hz}$. Similar frequency differences can
be obtained by slightly varying the geometry of the resonators or
the thickness of the deposited gold layer.\\
\begin{figure}[t]
\includegraphics [width=85mm] {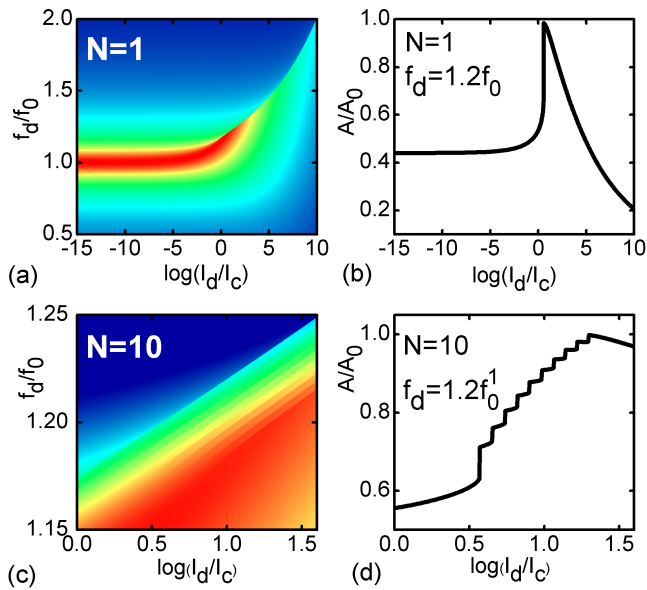}
\caption{Calculated amplitude responses. (a) Normalized amplitude
response of a single resonator driven at fixed frequency as a
function of the normalized drive amplitude $f_d / f_0$. (b) Cross
section at $f_d / f_0=1.2$. (c) and (d) show the amplitude
response for ten resonators.}
\end{figure}
\indent\indent To corroborate the experimental results, we have
calculated the characteristic driving amplitude of a nonlinear
resonator at a given driving frequency by solving the equation of
motion:
\begin{equation}\label{eqmotion}
M\ddot{y}+C\dot{y}+(K_t+K_b)y+K_3y^3 =BLI_d\cos(\omega t),
\end{equation}
where $M$ is the resonator effective mass, y the beam
displacement $C$ the damping constant, and $K_t$ and $K_b$
represent linear stiffness terms due to residual tension and
bending rigidity respectively. The nonlinear term $K_3$ is a
result of the elongation of the beam, $L$ is the beam length, $B$
the magnetic field strength and
$I_d$ the driving current.\\
\indent\indent For the beams in Fig.2, the average residual
tension $T_0=74\mathrm{\mu N}$ is found by comparing the
calculated stress-free resonance frequency to the measured
resonance frequency \cite{Bokaian89}. We calculated that
$K_t/K_b\approx85$, which indicates that the bending rigidity of
the resonators is insignificant and the linear resonance frequency
is merely determined by $M$ and $K_t$. The magnetomotive voltage
generated by the string-like linear resonator at resonance equals:
\begin{equation}\label{eqVEMF}
V_{EMF}=\frac{4}{\pi^4}\frac{\omega_1I_d B^2QL^3}{T_0}.
\end{equation}
With the experimental values $B=1.9$ T, and $Q=39$, and $L=200$
$\mathrm{\mu m}$, the  transduction at resonance equals
$V_{EMF}/I_d=0.8$ $\mathrm{V/A}$.\\
\indent\indent Instead of the frequency-dependent resonator
amplitude, we now calculate the resonator amplitude while varying
the drive amplitude at constant frequency $f_d$ to mimic the
experimental situation. Thus, out of two stable states, the
resonator always vibrates at the lowest amplitude. Figure 4(a)
shows the calculated response for a resonator similar to ones used
in the experiments. The frequency axis is normalized to the linear
resonance frequency $f_0$, and traces for the normalized drive
frequencies $f_d / f_0$ from $0.5$ to $2.0$ are combined in (a).
Red regions correspond to a high amplitude. The trace for $f_d
/f_0=1.2$ is shown in Fig.4(b). Sweeping parallel to the
horizontal axis, locations with multi-valued resonator amplitudes
are accessible at large drive amplitudes, when $f_d>f_0$. Figure
4(c) shows the simulated response of an array of $10$ resonators
described by Eq. \ref{eqmotion}, where a slight difference in
$K_t$ results in different characteristic amplitudes for each
resonator. We assume a difference in linear resonance frequency
for adjacent resonators equal to one tenth of the resonator
bandwidth. Panel (d) shows a cross section of the amplitude
response at $f_d /f_{0}^{1}=1.2$, where $f_{0}^{1}$ is the linear
resonance frequency of the slowest resonator. The phase response
(not shown) also displays steep transitions similar to the
amplitude, whose location can be determined with high accuracy
\cite{Greywall94}. The calculations thus reproduce the
experimental findings in detail.\\
\indent\indent In conclusion, we have demonstrated an efficient
way to operate and readout mechanical resonator arrays in
dissipative environments. The center frequencies of the linear
resonators can be spaced very closely as each resonator is marked
by its characteristic drive amplitude at a fixed frequency,
instead of the linear resonance frequency. As the exponential
relation between required bandwidth and the number of resonators
is circumvented, this technique may enable the construction of
high-throughput resonant sensor arrays.\\
\indent\indent The authors acknowledge financial support from
Koninklijke Philips NV (RWC-061-JR-05028) and from the Dutch
organizations FOM and NWO (VICI).


\begin{thebibliography}{14}
\expandafter\ifx\csname
natexlab\endcsname\relax\def\natexlab#1{#1}\fi
\expandafter\ifx\csname bibnamefont\endcsname\relax
  \def\bibnamefont#1{#1}\fi
\expandafter\ifx\csname bibfnamefont\endcsname\relax
  \def\bibfnamefont#1{#1}\fi
\expandafter\ifx\csname citenamefont\endcsname\relax
  \def\citenamefont#1{#1}\fi
\expandafter\ifx\csname url\endcsname\relax
  \def\url#1{\texttt{#1}}\fi
\expandafter\ifx\csname
urlprefix\endcsname\relax\def\urlprefix{URL }\fi
\providecommand{\bibinfo}[2]{#2}
\providecommand{\eprint}[2][]{\url{#2}}

\bibitem[{\citenamefont{Karabacak et~al.}(2006)\citenamefont{Karabacak, Kouh,
  Huang, and Ekinci}}]{Karabacak06}
\bibinfo{author}{\bibfnamefont{D.}~\bibnamefont{Karabacak}},
  \bibinfo{author}{\bibfnamefont{T.}~\bibnamefont{Kouh}},
  \bibinfo{author}{\bibfnamefont{C.~C.} \bibnamefont{Huang}}, \bibnamefont{and}
  \bibinfo{author}{\bibfnamefont{K.~L.} \bibnamefont{Ekinci}},
  \bibinfo{journal}{Appl. Phys. Lett.} \textbf{\bibinfo{volume}{88}},
  \bibinfo{pages}{19322} (\bibinfo{year}{2006}).

\bibitem[{\citenamefont{Li et~al.}(2007)\citenamefont{Li, Tang, and
  Roukes}}]{Li07}
\bibinfo{author}{\bibfnamefont{M.}~\bibnamefont{Li}},
  \bibinfo{author}{\bibfnamefont{H.~X.} \bibnamefont{Tang}}, \bibnamefont{and}
  \bibinfo{author}{\bibfnamefont{M.~L.} \bibnamefont{Roukes}},
  \bibinfo{journal}{Nat. Nanotechnol.} \textbf{\bibinfo{volume}{2}},
  \bibinfo{pages}{114} (\bibinfo{year}{2007}).

\bibitem[{\citenamefont{Dohn et~al.}(2006)\citenamefont{Dohn, Hansen, and
  Boisen}}]{Dohn06}
\bibinfo{author}{\bibfnamefont{S.}~\bibnamefont{Dohn}},
  \bibinfo{author}{\bibfnamefont{O.}~\bibnamefont{Hansen}}, \bibnamefont{and}
  \bibinfo{author}{\bibfnamefont{A.}~\bibnamefont{Boisen}},
  \bibinfo{journal}{Appl. Phys. Lett.} \textbf{\bibinfo{volume}{88}},
  \bibinfo{pages}{264104} (\bibinfo{year}{2006}).

\bibitem[{\citenamefont{Postma et~al.}(2005)\citenamefont{Postma, Kozinsky,
  Husain, and Roukes}}]{Postma05}
\bibinfo{author}{\bibfnamefont{H.~W.~C.} \bibnamefont{Postma}},
  \bibinfo{author}{\bibfnamefont{I.}~\bibnamefont{Kozinsky}},
  \bibinfo{author}{\bibfnamefont{A.}~\bibnamefont{Husain}}, \bibnamefont{and}
  \bibinfo{author}{\bibfnamefont{M.~L.} \bibnamefont{Roukes}},
  \bibinfo{journal}{Appl. Phys. Lett.} \textbf{\bibinfo{volume}{86}}
  (\bibinfo{year}{2005}).

\bibitem[{\citenamefont{Mertens et~al.}(2005)\citenamefont{Mertens, Alvarez,
  and Tamayo}}]{Mertens05}
\bibinfo{author}{\bibfnamefont{J.}~\bibnamefont{Mertens}},
  \bibinfo{author}{\bibfnamefont{M.}~\bibnamefont{Alvarez}}, \bibnamefont{and}
  \bibinfo{author}{\bibfnamefont{J.}~\bibnamefont{Tamayo}},
  \bibinfo{journal}{Appl. Phys. Lett.} \textbf{\bibinfo{volume}{87}}
  (\bibinfo{year}{2005}).

\bibitem[{\citenamefont{Yue et~al.}(2008)\citenamefont{Yue, Stachowiak, Lin,
  Datar, Cote, and Majumdar}}]{Yue08}
\bibinfo{author}{\bibfnamefont{M.}~\bibnamefont{Yue}},
  \bibinfo{author}{\bibfnamefont{J.~C.} \bibnamefont{Stachowiak}},
  \bibinfo{author}{\bibfnamefont{H.}~\bibnamefont{Lin}},
  \bibinfo{author}{\bibfnamefont{R.}~\bibnamefont{Datar}},
  \bibinfo{author}{\bibfnamefont{R.}~\bibnamefont{Cote}}, \bibnamefont{and}
  \bibinfo{author}{\bibfnamefont{A.}~\bibnamefont{Majumdar}},
  \bibinfo{journal}{Nano Lett.} \textbf{\bibinfo{volume}{8}},
  \bibinfo{pages}{520} (\bibinfo{year}{2008}).

\bibitem[{\citenamefont{Truitt et~al.}(2007)\citenamefont{Truitt, Hertzberg,
  Huang, Ekinci, and Schwab}}]{Ekinci07}
\bibinfo{author}{\bibfnamefont{P.~A.} \bibnamefont{Truitt}},
  \bibinfo{author}{\bibfnamefont{J.~B.} \bibnamefont{Hertzberg}},
  \bibinfo{author}{\bibfnamefont{C.~C.} \bibnamefont{Huang}},
  \bibinfo{author}{\bibfnamefont{K.~L.} \bibnamefont{Ekinci}},
  \bibnamefont{and} \bibinfo{author}{\bibfnamefont{K.~C.}
  \bibnamefont{Schwab}}, \bibinfo{journal}{Nano Lett.}
  \textbf{\bibinfo{volume}{7}} (\bibinfo{year}{2007}).

\bibitem[{\citenamefont{Van~Eysden and Sader}(2006)}]{Sader06JAP}
\bibinfo{author}{\bibfnamefont{C.~A.} \bibnamefont{Van~Eysden}}
  \bibnamefont{and} \bibinfo{author}{\bibfnamefont{J.~E.} \bibnamefont{Sader}},
  \bibinfo{journal}{J. Appl. Phys.} \textbf{\bibinfo{volume}{100}}
  (\bibinfo{year}{2006}).

\bibitem[{\citenamefont{Greywall}(2005)}]{Greywall05}
\bibinfo{author}{\bibfnamefont{D.~S.} \bibnamefont{Greywall}},
  \bibinfo{journal}{Meas. Sci. Technol.} \textbf{\bibinfo{volume}{16}},
  \bibinfo{pages}{2473–2482} (\bibinfo{year}{2005}).

\bibitem[{\citenamefont{Gardeniers et~al.}(1996)\citenamefont{Gardeniers,
  Tilmans, and Visser}}]{Gardeniers96}
\bibinfo{author}{\bibfnamefont{J.~G.~E.} \bibnamefont{Gardeniers}},
  \bibinfo{author}{\bibfnamefont{H.~A.~C.} \bibnamefont{Tilmans}},
  \bibnamefont{and} \bibinfo{author}{\bibfnamefont{C.~C.~G.}
  \bibnamefont{Visser}}, \bibinfo{journal}{J. Vac. Sci. Technol. A}
  \textbf{\bibinfo{volume}{14}}, \bibinfo{pages}{2879} (\bibinfo{year}{1996}).

\bibitem[{\citenamefont{Halbach}(1983)}]{Halbach83}
\bibinfo{author}{\bibfnamefont{K.}~\bibnamefont{Halbach}},
  \bibinfo{journal}{IEEE Trans. Nucl. Sci.} \textbf{\bibinfo{volume}{NS-30}},
  \bibinfo{pages}{3323} (\bibinfo{year}{1983}).

\bibitem[{\citenamefont{Cleland and Roukes}(2002)}]{Cleland02}
\bibinfo{author}{\bibfnamefont{A.~N.} \bibnamefont{Cleland}} \bibnamefont{and}
  \bibinfo{author}{\bibfnamefont{M.~L.} \bibnamefont{Roukes}},
  \bibinfo{journal}{J. Appl. Phys.} \textbf{\bibinfo{volume}{92}},
  \bibinfo{pages}{2758} (\bibinfo{year}{2002}).

\bibitem[{\citenamefont{Bokaian}(1990)}]{Bokaian89}
\bibinfo{author}{\bibfnamefont{A.}~\bibnamefont{Bokaian}}, \bibinfo{journal}{J.
  Sound Vib.} \textbf{\bibinfo{volume}{142}}, \bibinfo{pages}{481}
  (\bibinfo{year}{1990}).

\bibitem[{\citenamefont{Greywall et~al.}(1994)\citenamefont{Greywall, Yurke,
  Busch, Pargellis, and Willet}}]{Greywall94}
\bibinfo{author}{\bibfnamefont{D.~S.} \bibnamefont{Greywall}},
  \bibinfo{author}{\bibfnamefont{B.}~\bibnamefont{Yurke}},
  \bibinfo{author}{\bibfnamefont{P.~A.} \bibnamefont{Busch}},
  \bibinfo{author}{\bibfnamefont{A.~N.} \bibnamefont{Pargellis}},
  \bibnamefont{and} \bibinfo{author}{\bibfnamefont{R.~L.}
  \bibnamefont{Willet}}, \bibinfo{journal}{Phys. Rev. Lett.}
  \textbf{\bibinfo{volume}{72}}, \bibinfo{pages}{2992} (\bibinfo{year}{1994}).
\end{thebibliography}
\end{document}